\documentclass[10pt]{article}

\usepackage[margin=1in]{geometry}
\usepackage{graphicx}
\usepackage{amsmath,amssymb,braket}
\usepackage{cite}
\usepackage{authblk}
\usepackage{hyperref}
\usepackage{xcolor}
\usepackage{lineno}

\title{Latent symmetry in a minimal non-Hermitian trimer}

\author[1]{P. A. Brand\~ao\thanks{paulo.brandao@fis.ufal.br}}
\affil[1]{Instituto de F\'isica, Universidade Federal de Alagoas, Av. Lourival Melo Mota, S/N, Tabuleiro do Martins, 57072-970, Macei\'o, Alagoas, Brasil. }

\date{} 

\begin{document}
	
	\maketitle
	
	\begin{abstract}
		We study a minimal non-Hermitian trimer with latent symmetry formed by a cospectral pair of sites embedded in a three-site network with nonreciprocal couplings. We show that   cospectrality provides a structural latent-symmetry constraint, whereas exact dark-state decoupling requires an additional algebraic matching condition among the couplings. For a dark-state-compatible representative of this cospectral class, the model admits an exact decomposition into dark and bright sectors: the dark mode is spectrally isolated and retains a complex eigenvalue, while the bright sector reduces to an effective non-Hermitian dimer.  For a suitable choice of parameters, this reduced subsystem becomes $\mathcal{PT}$-symmetric and exhibits partial spectral reality, with two real eigenvalues coexisting with the complex dark eigenvalue. At the critical point, the bright sector hosts an embedded second-order exceptional point, which renders the full trimer defective and gives rise to the characteristic Jordan-block dynamics. These results establish the non-Hermitian trimer as a minimal analytically solvable setting in which latent symmetry, sector-resolved $\mathcal{PT}$ symmetry, and exceptional-point physics naturally coexist.
	\end{abstract}

	\section{Introduction}
	
	Non-Hermitian Hamiltonians have become a central tool for describing effective dynamics in open and wave-based systems with gain, loss, and asymmetric mode conversion. In particular, parity-time ($\mathcal{PT}$) symmetry showed that non-Hermitian operators may still display entirely real spectra within a finite parameter window, followed by a symmetry-breaking transition in which eigenvalues become complex \cite{Bender1998,ElGanainy2018}. Optical settings have played a decisive role in this development because coupled waveguides, resonators, and synthetic lattices provide a flexible platform in which gain, loss, and nonreciprocal couplings can be engineered and probed directly \cite{Guo2009,Ruter2010,ElGanainy2018,Ozdemir2019,Miri2019}. A hallmark of such systems is the presence of exceptional points (EPs), non-Hermitian degeneracies at which both eigenvalues and eigenvectors coalesce, leading to defective Hamiltonians and anomalous dynamical responses \cite{Heiss2012,Ozdemir2019,Miri2019}.
	
	In parallel, recent work has shown that spectral graph concepts such as cospectrality can generate a more hidden form of order in networks. Two sites are cospectral when the corresponding vertex-deleted subgraphs have the same characteristic polynomial \cite{Schwenk1973,Kempton2020}. This notion underlies latent symmetry, a generalized symmetry that is not necessarily visible in the full graph but emerges after suitable spectral reduction \cite{Smith2019}. In photonic lattices, latent symmetry was recently shown to constrain eigenmodes and transport in a robust and experimentally accessible way, opening a new route to mode engineering beyond conventional geometric symmetries \cite{Himmel2025,Himmel2026}.
	
	These two directions suggest a natural question: how does latent symmetry manifest itself in a genuinely non-Hermitian setting? More specifically, can a minimal non-Hermitian network combine cospectrality, dark-state formation, and $\mathcal{PT}$-type spectral transitions in a single analytically tractable model? Addressing this question is appealing both conceptually and practically. Conceptually, latent symmetry is formulated through spectral constraints, whereas non-Hermitian physics is governed by complex spectra, defective eigenspaces, and sector-dependent stability. Practically, minimal non-Hermitian trimers already serve as canonical models for coupled photonic structures and provide the simplest setting in which one may embed a nontrivial spectator mode alongside a reduced dimer dynamics.
	
	Nonreciprocal couplings of the kind considered here are no longer purely theoretical. In chiral quantum optics, spin-momentum locking in guided modes gives rise to propagation-direction-dependent emission, scattering, and absorption, thereby providing a natural route to effective nonreciprocal interactions in reduced models \cite{Lodahl2017}. Experimentally, such chiral interfaces have been demonstrated with cold atoms coupled to optical nanofibres, where spontaneous emission into the two counter-propagating guided modes can be tuned from symmetric to strongly asymmetric \cite{Mitsch2014}, with semiconductor quantum dots embedded in nanophotonic waveguides, where near-unity directionality and deterministic chiral photon-emitter coupling have been achieved \cite{Sollner2015,Coles2016}, and with superconducting circuits, where on-demand directional microwave photon emission and strong unidirectional coupling have recently been reported \cite{Kannan2023,Joshi2023}. Related cascaded radiative dynamics have also been observed in waveguide-coupled atomic platforms \cite{Liedl2024}. In this sense, the asymmetric couplings employed in our minimal trimer may be viewed as an effective reduced description of experimentally accessible chiral and cascaded photonic architectures.
	
	Several non-Hermitian trimers have been investigated in the context of $\mathcal{PT}$ symmetry and related spectral transitions. Early work on $\mathcal{PT}$-symmetric oligomers established the trimer as one of the minimal few-site platforms displaying non-Hermitian phase transitions, nonlinear stationary states, and nontrivial stability properties \cite{Li2011}. This picture was later refined in a systematic study of the $\mathcal{PT}$-symmetric trimer, where bifurcations, ghost states, and the associated unstable dynamics were analyzed in detail \cite{Li2013}. Beyond globally $\mathcal{PT}$-symmetric gain-neutral-loss arrangements, it was subsequently shown that non-Hermitian trimers may also exhibit entirely real spectra in the absence of explicit spatial $\mathcal{PT}$ symmetry, highlighting the broader role of pseudo-Hermitian structures in few-mode systems \cite{Suchkov2016}. In a different direction, trimers have also been used as parent platforms for the emergence of effective non-Hermitian dimers, for instance through adiabatic elimination in coupled waveguides \cite{Sharaf2018}. More recently, triangular bosonic trimers with complex couplings have revealed the interplay between $\mathcal{PT}$ symmetry and chiral current circulation, further emphasizing the richness of three-site non-Hermitian geometries \cite{Downing2020}. In this context, the present work identifies a distinct minimal mechanism: rather than imposing a global $\mathcal{PT}$ symmetry on the full trimer, we exploit latent symmetry to generate an exact dark--bright decomposition, in which the bright invariant sector reduces to an effective $\mathcal{PT}$-symmetric dimer while the dark mode remains spectrally isolated.

	\section{Trimer model and cospectrality}
	
\begin{figure}[ht]
		\centering
		\includegraphics[width=.4\linewidth]{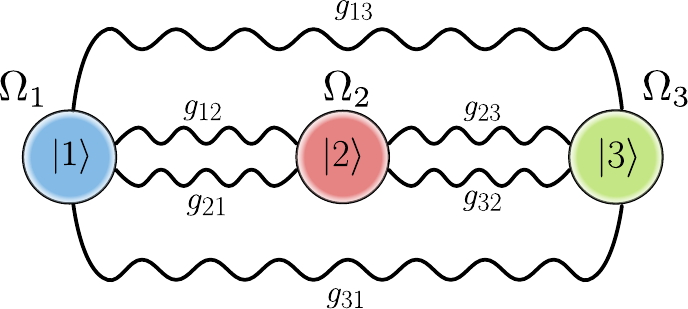}
		\caption{Minimal non-Hermitian trimer formed by three coupled sites, $\ket{1}$, $\ket{2}$, and $\ket{3}$, with complex onsite energies $\Omega_j = \omega_j+i\gamma_j $. The pair ($\ket{1}$, $\ket{2}$) becomes cospectral when \(\Omega_1=\Omega_2\) and \(g_{13}g_{31}=g_{23}g_{32}\), giving rise to a latent symmetry, while site \(|3\rangle\) plays the role of a singlet site.}
		\label{fig1}
	\end{figure}

	Consider a physical configuration of three connected sites $\ket{1}$, $\ket{2}$ and $\ket{3}$ having complex energies $\Omega_j = \omega_j + i\gamma_j$ ($j = 1,2,3)$ with $\omega_j,\gamma_j\in\mathbb{R}$, as displayed in Fig. \ref{fig1}. The effective non-Hermitian Hamiltonian describing this configuration can be written in a tight-binding form as
	\begin{equation}
		H = \sum_{j=1}^3\Omega_j \ket{j}\bra{j} + \sum_{i=1}^3\sum_{j\neq i}^3g_{ij}\ket{i}\bra{j},
	\end{equation}
	where $g_{ij}$ are the real-valued coupling constants. In the basis $\{ \ket{1},\ket{2},\ket{3}  \}$, the Hamiltonian can be represented by the complex matrix
	\begin{equation}\label{originalM}
		H = \begin{pmatrix}
			\omega_1 + i\gamma_1 & g_{12} & g_{13} \\
			g_{21} & \omega_2 + i\gamma_2 & g_{23} \\
			g_{31} & g_{32} & \omega_3 + i\gamma_3
		\end{pmatrix}.
	\end{equation}
	
	The latent symmetry of a network is constructed in terms of its cospectrality. Two sites $\ket{i}$ and $\ket{j}$ are said to be \emph{cospectral} if the vertex-deleted subgraphs, $H \backslash \ket{i}$ and $H \backslash \ket{j}$, have the same characteristic polynomial \cite{Schwenk1973,Smith2019,Kempton2020}. Let us use this concept to develop a set of conditions such that sites $\ket{1}$ and $\ket{2}$ become cospectral. The subgraphs $H \backslash \ket{1}$ and $H \backslash\ket{2}$ obtained by removing sites $\ket{1}$ and $\ket{2}$, respectively, are represented by the matrices
	\begin{align}
		H \backslash\ket{1}  &= \begin{pmatrix}
			\omega_2 + i\gamma_2 &  g_{23} \\
			g_{32} & \omega_3 + i\gamma_3
		\end{pmatrix}, \\ 
		H \backslash\ket{2} &=  \begin{pmatrix}
			\omega_1 + i\gamma_1 &  g_{13} \\
			g_{31} & \omega_3 + i\gamma_3
		\end{pmatrix}.
	\end{align} 
	It is a simple exercise to demonstrate that the eigenvalues of the above matrices are equal if and only if
	\begin{equation}\label{conditions}
		\omega_1 = \omega_2 = \omega, \, \gamma_1 = \gamma_2 = \gamma \quad \text{and} \quad g_{13}g_{31} = g_{23}g_{32}.
	\end{equation}
	The above conditions on the cospectrality between $\ket{1}$ and $\ket{2}$ imply that latent symmetry does not require gain-loss balance between sites 1 and 2. It only requires equal complex onsite potentials and a product constraint on the couplings. Thus, one has additional freedom regarding the coupling constants, since the last condition on \eqref{conditions} does not imply that the couplings $g_{ij}$ need to be symmetric, in the sense that $g_{ij} = g_{ji}$.
	
	Suppose that conditions given by \eqref{conditions} are satisfied for some particular set of parameters and that the three eigenvalues of $H$ are nondegenerate. Then, the sites $\ket{1}$ and $\ket{2}$ exhibit a latent symmetry and the trimer is said to be latent symmetric. It is possible for a latent symmetric structure to feature singlet sites. They are defined as the set of sites whose distances (in the sense of a graph metric) to a pair of sites exhibiting a latent symmetry are equal. In our trimer model, since there is only one extra site, $\ket{3}$, it is a singlet site since it has the same distance to $\ket{1}$ and $\ket{2}$. What is interesting about singlet sites is that depending on the initial excitation on the pair ($\ket{1}$,$\ket{2}$), the dynamics of $\ket{3}$ is trivial \cite{Himmel2025}.

	It is important to distinguish the structural meaning of cospectrality from the dynamical condition for dark-state decoupling. In the present non-Hermitian setting, cospectrality means that the two latent sites are indistinguishable at the level of the spectra of the corresponding vertex-deleted Hamiltonians. This is a structural latent-symmetry constraint. However, cospectrality alone does not imply the existence of a right eigenstate with zero amplitude on the singlet site. The latter requires an additional destructive-interference condition among the couplings connecting the latent pair to the singlet site, together with a compatibility condition involving the internal couplings of the latent pair. These conditions are derived in Appendix A. The parametrized Hamiltonian introduced below should therefore be understood as a dark-state-compatible representative of the cospectral class.

	We now introduce a particular representative of this cospectral class which also 
	satisfies the dark-state matching condition derived in Appendix A. This additional 
	condition is not implied by cospectrality alone. It is imposed here in order to 
	obtain an exact dark-bright decomposition in a minimal analytically tractable 
	model. We write 
	\begin{equation}\label{mainM}
		H(\chi) = \begin{pmatrix}
			\Omega & \mu e^{2\chi} & \kappa e^{\chi} \\
			\mu e^{-2\chi} & \Omega & \kappa e^{-\chi} \\
			\kappa e^{-\chi} & \kappa e^{\chi} & \Omega_3
		\end{pmatrix},
	\end{equation}
	where $\Omega = \omega + i\gamma$, $\mu>0$, $\kappa >0$.   Indeed, for this parametrization one has $g_{13}g_{31}=g_{23}g_{32}=\kappa^2$, so that the pair $(\ket{1},\ket{2})$ is cospectral. In addition, $g_{12}/g_{21}=e^{4\chi} = \left(g_{32}/g_{31}\right)^2$, which is precisely the dark-state matching condition. Therefore, $H(\chi)$ is	not the most general cospectral trimer, but a dark-state-compatible representative of the cospectral class.   This form is more convenient and we can study the entire dynamics in terms of $\Omega$, $\mu$, $\kappa$ and $\chi$. Notice that \eqref{mainM} is related to \eqref{originalM} (with the conditions discussed above) by the similarity transformation $H(\chi) \rightarrow DHD^{-1}$, where $D = \text{diag}(e^{\chi},e^{-\chi}, 1)$, and so $\chi$ cannot change the eigenvalue structure of \eqref{originalM}.

	\section{Results and discussion}

	We now demonstrate that   the dynamics of the model represented by Eq. \eqref{mainM} can be described in terms of independent   dark and bright sectors. The bright sector can exhibit a $\mathcal{PT}$-symmetric phase transition with real eigenvalues while the dark sector remains in the ``broken'' phase with a single complex eigenvalue. The easiest way to see this is by noting that $\ket{D} = e^{\chi}\ket{1} - e^{-\chi}\ket{2}$ is an eigenvector of $H(\chi)$ with eigenvalue $\lambda_0 = \Omega - \mu = \omega - \mu + i\gamma$. This implies that site $\ket{3}$ is invisible to the dynamics generated by the invariant subspace span$\{ \ket{D} \}$, which is the dark sector of the trimer. If the initial state is given by $\ket{\psi(0)} = \ket{D}$, then $\ket{\psi(t)} = e^{-iHt} \ket{\psi(0)} = e^{-i(\omega - \mu)t}e^{\gamma t} (e^{\chi}\ket{1} - e^{-\chi}\ket{2})$. Thus, the local occupations $P_j = |\braket{j | \psi(t)}|^2$ of sites $j = 1,2,3$ are given by $P_1 = e^{2\gamma t}e^{2\chi}$, $P_2 = e^{2\gamma t}e^{-2\chi}$ and $P_3 = 0$.   For the dark-state-compatible latent-symmetric family in Eq. \eqref{mainM}, an excitation prepared in the antisymmetric combination of the latent sites remains confined to the one-dimensional invariant subspace spanned by $\ket{D}$, and the singlet-site occupation $P_3$ vanishes identically.

\begin{figure}[ht]
		\centering
		\includegraphics[width=.5\linewidth]{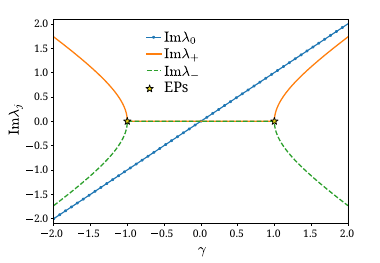}
		\caption{Imaginary parts of the eigenvalues $\lambda_0$, $\lambda_+$ and $\lambda_-$ as a function of $\gamma$. The EPs are located at $\gamma_c = \pm 1$ for the set of parameters $\omega = 0$, $\mu = 1$ and $\kappa = 1/\sqrt{2}$. }
		\label{fig2}
	\end{figure}

	On the other hand, for the state $\ket{B} = (e^{\chi} \ket{1} + e^{-\chi}\ket{2})/\sqrt{2}$ one has $H(\chi)\ket{B} = (\Omega + \mu)\ket{B} + \sqrt{2}\kappa \ket{3}$ and $H(\chi) \ket{3} = \sqrt{2}\kappa \ket{B} + \Omega_3\ket{3}$. Therefore, the set span$\{  \ket{B},\ket{3}  \}$ generates another invariant subspace which is the bright sector. In this basis, the dynamics is described by the $2\times 2$ matrix
	\begin{equation}\label{subblock}
		H(\chi)_{B} = \begin{pmatrix}
			\Omega + \mu & \sqrt{2}\kappa \\
			\sqrt{2}\kappa & \Omega_3
		\end{pmatrix},
	\end{equation}
	which has eigenvalues
	\begin{equation}\label{eigplusminus}
		\lambda_{\pm}  = \frac{\Omega + \Omega_3 + \mu \pm \Delta }{2}, 
	\end{equation}
	where $\Delta = [(\Omega + \mu - \Omega_3)^2 + 8\kappa^2]^{1/2}$. Can this bright sector generate stable evolution in the sense of having nondivergent behavior for $P_j$? Since $\Omega = \omega + i\gamma$ and $\Omega_3 = \omega_3 + i\gamma_3$, we see from \eqref{eigplusminus} that in order to have $\lambda_{\pm} \in \mathbb{R}$ it is necessary that
	\begin{equation}\label{conditionreal}
		\gamma_3 = -\gamma \quad \text{and} \quad \omega_3 = \omega + \mu.
	\end{equation}
	This condition imposes that dissipation present in mode $\ket{3}$ must be of opposite sign compared to that in modes $\ket{1(2)}$ and also that there must be a frequency mismatch between $\ket{3}$ and $\ket{1(2)}$. We assume this to be the case in what follows. The matrix given by \eqref{subblock} can now be written as
	\begin{equation}\label{matrixPT}
		H(\chi)_B = \begin{pmatrix}
			\omega + \mu +i\gamma & \sqrt{2}\kappa \\
			\sqrt{2}\kappa & \omega + \mu -i\gamma
		\end{pmatrix},
	\end{equation}
	which represents a $\mathcal{PT}$-symmetric dimer \cite{huerta2016revisiting, jin2018parity, naikoo2021pt}. From this, we conclude that the dynamics generated from a symmetric excitation in the latent sites $\ket{1}$ and $\ket{2}$ can only connect the site $\ket{3}$ with the bright mode $\ket{B}$. The discriminant now reads $\Delta = [(\Omega + \mu - \Omega_3)^2 + 8\kappa^2]^{1/2} = (-4\gamma^2 + 8\kappa^2)^{1/2}$ which shows that the evolution is stable if $\gamma$ is selected in the range $\gamma \in (-\sqrt{2}\kappa, \sqrt{2}\kappa)$. The critical point $\gamma_c = \sqrt{2}\kappa$ corresponds to an EP of second order of the full trimer associated with the bright invariant subspace. Thus, in the special case where $\gamma\in (-\sqrt{2}\kappa, \sqrt{2}\kappa)$, we have the real eigenvalues $\lambda_{\pm}^r = \omega + \mu \pm (2\kappa^2 -\gamma^2)^{1/2}$. The imaginary parts of $\lambda_j$ are displayed in Fig. \ref{fig2}.
	
	In the case of symmetric initial excitation in the latent sites, $\ket{\psi(0)} = \ket{B}$, and in the absence of EPs, the evolved state is given by
	\begin{equation}
		\ket{\psi(t)} = e^{-iat} \Bigg[ \frac{\alpha(t)}{\sqrt{2}}(e^{\chi}\ket{1} + e^{-\chi}\ket{2}) + \beta(t)\ket{3}  \Bigg],
	\end{equation}
	where $\alpha(t) = \cos(\eta t) - (i\delta/\eta)\sin(\eta t)$, $\beta(t) = -i\sqrt{2}\kappa\sin(\eta t)/\eta$, $\eta = \Delta/2$, $\delta = (\Omega + \mu -\Omega_3)/2$ and $a = (\Omega + \mu + \Omega_3)/2$. The local occupations are now
	\begin{equation}\label{p1bright}
		P_1 = \frac{e^{2\chi}|\alpha(t)|^2}{2},
	\end{equation}
	\begin{equation}\label{p2bright}
		P_2 = \frac{e^{-2\chi}|\alpha(t)|^2}{2},
	\end{equation}
	\begin{equation}\label{p3bright}
		P_3 = |\beta(t)|^2.
	\end{equation}
	We thus observe a coherent and bounded energy exchange between the latent sites $\ket{1}$ and $\ket{2}$ and the singlet mode $\ket{3}$ even though the spectrum of the trimer has a complex eigenvalue $\lambda_0$.
	
	The cospectral analysis   and dark-state formation   has revealed an EP living in the bright sector of the trimer when the condition $\gamma = \gamma_c = \sqrt{2}\kappa$ is satisfied. In this case, the matrix given by \eqref{matrixPT} can be written as $H(\chi)_B = (\omega + \mu )I + N$, where $I$ is the $2x2$ identity matrix and 
	\begin{equation}
		N = \gamma_c\begin{pmatrix}
			i & 1 \\
			1 & -i
		\end{pmatrix}
	\end{equation}
	is nilpotent to second-order: $N^2 = 0$. The eigenvalues $\lambda_{\pm}$ given by \eqref{eigplusminus} become equal to $\lambda_c = \omega + \mu$ at $\gamma = \gamma_c$ and the (right) eigenvectors coalesce to $\ket{\phi_c} = \ket{B} - i\ket{3}$. If the initial state is given by the bright state $\ket{\psi(0)} = \ket{B}$, then $\ket{\psi(t)} = e^{-i(\omega + \mu)t}e^{-iNt}\ket{B} = e^{-i(\omega + \mu)t}(\ket{B} - it N\ket{B}) = e^{-i(\omega + \mu)t}[(1 + \gamma_c t)\ket{B} - i\gamma_c t \ket{3}]$. At the EP, the local occupations grow polynomially as $P_1 = e^{2\chi}(1+\gamma_c t)^2/2$, $P_2 = e^{-2\chi}(1+\gamma_c t)^2/2$ and $P_3 = \gamma_c^2t^2$. 
	
	\begin{figure}[ht]
		\centering
		\includegraphics[width=.6\linewidth]{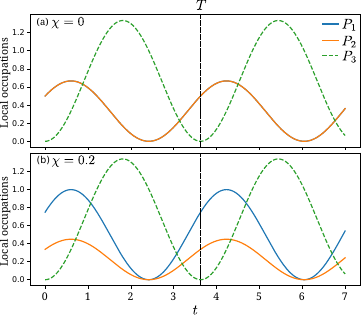}
		\caption{Dynamics in the bright sector ($\ket{\psi(0)} = \ket{B})$ below the $\mathcal{PT}$ phase transition point. Local occupations $P_j = |\braket{j | \psi(t)}|^2$  for (a) $\chi = 0$ and (b) $\chi = 1/5$. Other parameters are the same as in Fig. \ref{fig2} and $\gamma = 1/2$.}
		\label{fig3}
	\end{figure}
	
	Since the bright sector holds the stable dynamics, we now give numerical examples by solving $\ket{\psi(t)} = e^{-iH(\chi)t}\ket{\psi(0)}$, where $H(\chi)$ is given by \eqref{mainM} and $\ket{\psi(0)} = \ket{B}$ is the initial bright state of the trimer. Fig. \ref{fig3} plots the local occupations $P_j(t)$ for (a) $\chi = 0$ and (b) $\chi = 1/5$. In the first case, there is a periodic energy exchange between $(\ket{1},\ket{2}) \leftrightarrow \ket{3}$. The energy is equally distributed between the two cospectral sites $\ket{1}$ and $\ket{2}$ during evolution. By changing the deformation parameter $\chi$, this energy balance can be lifted, as shown in part (b) of the same figure. Thus, one can control the amount of energy each site receives by tuning $\chi$ appropriately. This fact is evident from the form of \eqref{p1bright}, \eqref{p2bright} and \eqref{p3bright}, which shows that increasing $\chi$ has the effect of redistributing more energy into $\ket{1}$ while at the same time decreasing the energy content in site $\ket{2}$. Notice that $P_3$ is stable and independent of $\chi$, as shown by the dashed-dot green lines in the figure. For suitable values of $\chi$ one can in fact generate an effective dynamics between $\ket{1}$ and $\ket{3}$ in which they exchange the same amount of energy much larger than the maximum value of $P_3$.

	\begin{figure}[!htbp]
		\centering
		\includegraphics[width=.5\linewidth]{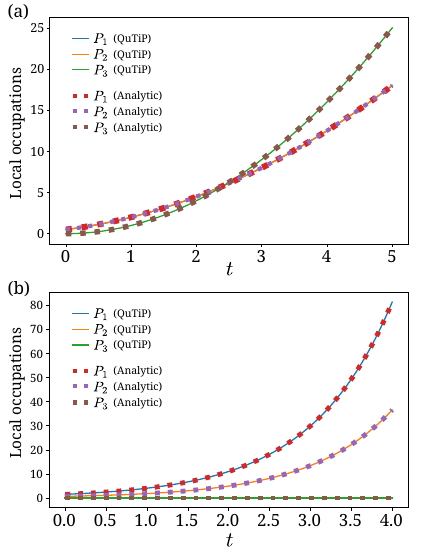}
		\caption{  (a) Dynamics of the critical bright-sector at the embedded EP, obtained for	$\ket{\psi(0)}=\ket{B}$, $\gamma=\gamma_c=\sqrt{2}\kappa=1$, $\omega=0$, $\mu=1$, $\kappa=1/\sqrt{2}$ and $\chi=0$. The local occupations display the polynomial growth expected from the Jordan-block structure, with $P_1(t) = P_2(t) = (1+\gamma_c t)^2/2$ and $P_3(t) = (\gamma_ct)^2$. (b) Dark-sector dynamics for $\ket{\psi(0)} = \ket{D}$, with $\gamma=1/2$ and $\chi=0.2$. The occupation of the singlet site remains identically zero, $P_3(t) = 0$, confirming the complete decoupling of the dark state from site $\ket{3}$. Solid curves show the numerical evolution under	the full trimer Hamiltonian, by using the QuTiP package in python \cite{lambert2026qutip}, while dotted curves show the corresponding analytical predictions.  }
		\label{fig4}
	\end{figure}
	 
	To complement the analytical results discussed above, we now provide direct numerical verifications of the invariant-sector dynamics by evolving the full trimer Hamiltonian. The results are shown in Fig. \ref{fig4}. In Fig.\ref{fig4}(a), the system is initialized in the bright state $\ket{B}$ and the gain-loss parameter is 
	fixed at the critical value $\gamma = \gamma_c = \sqrt{2}\kappa $. The local occupations $P_j$ display the expected polynomial growth, confirming the Jordan-block dynamics associated with the embedded second-order EP. Fig. \ref{fig4}(b) shows the complementary dark-sector evolution for the initial condition $\ket{\psi(0)} = \ket{D}$. In this case, the occupation $P_3$ of the singlet site remains identically zero during the whole evolution, while the occupations $P_1$ and $P_2$ of the two latent sites evolve according to the overall 
	non-Hermitian factor $e^{2\gamma t}$. This numerical result confirms that the dark state is completely decoupled from site $\ket{3}$, as predicted by the one-dimensional invariant subspace $\mathrm{span}\{\ket{D} \}$.

	\section{Conclusions}
	
	In conclusion, we have introduced a minimal non-Hermitian trimer in which a cospectral pair of sites gives rise to latent symmetry and an exact decomposition into dark and bright sectors   has been found.   The dark sector is formed by a spectrally isolated mode that remains decoupled from the singlet site, whereas the bright sector reduces exactly to an effective non-Hermitian dimer. For a suitable choice of onsite parameters, this reduced subsystem becomes $\mathcal{PT}$-symmetric and displays a regime of partial spectral reality, with two real eigenvalues coexisting with a complex dark eigenvalue. At the critical point, the bright sector hosts a second-order EP embedded in the full trimer, leading to the characteristic polynomial-in-time dynamics associated with a Jordan block. Our numerical examples further show that, below the $\mathcal{PT}$-transition threshold, the deformation parameter $\chi$ provides a simple way to redistribute the local occupations while preserving bounded oscillatory exchange in the bright sector. These results establish the trimer studied here as a minimal analytically solvable platform in which latent symmetry, non-Hermitian mode selection, and exceptional-point physics coexist, and they suggest a useful route for mode engineering in photonic structures with asymmetric couplings.

	\appendix
	
	\section{Generic dark-state condition}

	In this appendix, we derive the algebraic conditions for the existence of a right dark state supported only on the latent pair. Consider the generic non-Hermitian trimer
	\begin{equation}
	H =
	\begin{pmatrix}
		\Omega_1 & g_{12} & g_{13}\\
		g_{21} & \Omega_2 & g_{23}\\
		g_{31} & g_{32} & \Omega_3
	\end{pmatrix},
	\end{equation}	
	where $\Omega_j=\omega_j+i\gamma_j$. We now look for an unnormalized right 
	eigenstate of the form $\ket{D} =  \ket{1} + \lambda \ket{2}$, with vanishing amplitude on site $\ket{3}$. Acting with $H$ gives
	\begin{equation}
	H\ket{D} = (\Omega_1 + \lambda g_{12}) \ket{1} + (g_{21} + \lambda \Omega_2)\ket{2}	+
	(g_{31}+\lambda g_{32})\ket{3}.
	\end{equation}	
	For $\ket{D}$ to be a dark state with respect to the singlet site, the 
	coefficient of \(|3\rangle\) must vanish: $g_{31}+\lambda g_{32}=0$. Assuming $g_{32}\neq 0$, this gives
	\begin{equation}
	\lambda=-\frac{g_{31}}{g_{32}}.
	\end{equation}
	In addition, $\ket{D}$ must be a right eigenvector of $H$. Therefore, $g_{21} + \lambda\Omega_2	=\lambda(\Omega_1 + \lambda g_{12})$ or $g_{21} + \lambda(\Omega_2-\Omega_1)-\lambda^2 g_{12}=0$.	Substituting $\lambda = -g_{31}/g_{32}$, one obtains the general dark-state	condition
	\begin{equation}
	g_{21}g_{32}^2 + (\Omega_1-\Omega_2) g_{31}g_{32} - g_{12}g_{31}^2 = 0.
	\end{equation}
	
	For the cospectral case considered in the main text, $\Omega_1 = \Omega_2$, this 
	reduces to $g_{21}g_{32}^2 = g_{12}g_{31}^2$ or, equivalently, $$\frac{g_{12}}{g_{21}}=
	\left(\frac{g_{32}}{g_{31}}\right)^2.$$ Thus, cospectrality alone is not sufficient to guarantee the existence of the dark state. It must be supplemented by this internal-coupling matching condition. The corresponding right dark state can be written as $\ket{D} = g_{32}\ket{1} - g_{31}\ket{2}$, up to an arbitrary normalization factor. Its eigenvalue is $$\lambda_D = \Omega_1-g_{31}g_{12}/g_{32}.$$ For the parametrized Hamiltonian in Eq. \eqref{mainM}, one has $g_{31} = \kappa e^{-\chi}$, $g_{32} = \kappa e^{\chi}$,	and therefore $$\ket{D} \propto e^\chi \ket{1} - e^{-\chi}\ket{2}.$$ Moreover, $$\frac{g_{12}}{g_{21}} =	\frac{\mu e^{2\chi}}{\mu e^{-2\chi}}	= e^{4\chi}	= \left(\frac{g_{32}}{g_{31}}\right)^2.$$ Hence, the family $H(\chi)$ satisfies both the cospectrality condition and the additional dark-state matching condition.

	\subsection*{Funding}
	This work was supported by the Brazilian agency CNPq (Conselho Nacional de Desenvolvimento Cient\'ifico e Tecnol\'ogico).

	\subsection*{Competing interests}
	The authors declare that they have no competing interests.

	\bibliographystyle{unsrt}
	\bibliography{references}
	
\end{document}